\documentclass[a4paper,12pt]{article}
\usepackage{amssymb}
\usepackage{bm}
\newcommand{\Integer}{\mathbb Z}

\newfont{\bg}{cmr10 scaled\magstep3}
\newcommand{\Bzero}{\smash{\lower1ex\hbox{\bg 0}}}
\newcommand{\bra}[1]{\langle\,#1\,|}
\newcommand{\ket}[1]{|\,#1\,\rangle}

\newcommand{\Nbra}[1]{~_N(\,#1\,|}
\newcommand{\Nket}[1]{|\,#1\,)_N}
\newcommand{\Nbracket}[2]{~_N(\,#1\,|\,#2\,)_N}
\newcommand{\nn}{\nonumber\\}

\begin{document}
\title{Representation of $SU(\infty)$ Algebra for Matrix Models}
\author{{\sc Naofumi Kitsunezaki}\thanks{Email address:
kitsune@eken.phys.nagoya-u.ac.jp}~~and
{\sc Shozo Uehara}\thanks{Email address:
uehara@eken.phys.nagoya-u.ac.jp}\\
{\it Department of Physics, Nagoya University,}\\
{\it Chikusa-ku, Nagoya 464-8602, Japan}}
\date{}
\maketitle
\vspace{-80mm}
\begin{flushright}
	DPNU-02-18\\
	hep-th/0206157\\
	June 2002
\end{flushright}
\vspace{50mm}
\begin{abstract}
We investigate how the matrix representation of $SU(N)$ algebra
approaches that of the Poisson algebra in the large $N$ limit.
In the adjoint representation, the $(N^2-1)\times(N^2-1)$ matrices
of the $SU(N)$ generators go to those of the Poisson algebra in the
large $N$ limit.
However, it is not the case for the $N\times N$ matrices in the
fundamental representation.
\end{abstract}
\section{Introduction}
In these years, matrix models have been studied to investigate string
theories \cite{BFSS}-\cite{AIKO}.
It is important to clarify the limit of the size of matrices
going to infinity to define those theories. We also expect that
there are such field theories that are properly regularized by
matrices and hence they are useful to study string theories since we
know a lot of techniques to examine field theories.
It was pointed out long ago that the generators of $SU(N)$
approach those of the Poisson algebra when $N$ becomes large, and the
commutators of the $SU(N)$ algebra are close to the Poisson bracket
\cite{Hoppe}-\cite{hoppesch}.
Recently it was argued in ref.\cite{zunger} that the group of
area-preserving diffeomorphisms of any Riemann surface, or connected,
compact and orientable world-sheet, is equivalent to $SU(\infty)$.
This implies that matrix models could be formulated on the world-sheet
with the area-preserving diffeomorphisms.
On the other hand, there are some investigations that the bosonic part
of the IIB matrix model, which is naively a matrix regularization of
the Schild model \cite{Schild:vq}, do not agree with the Schild model
in the large $N$ limit \cite{KU,ANO}.

In this paper, we study how the large $N$ limit of the generators of
$SU(N)$ go to those of the Poisson algebra in some representations.
In the next section, we show that in the adjoint representation, all
the matrix elements of the $SU(N)$ generators approach to those of the
Poisson algebra in the large $N$ limit.
On the other hand, fundamental representation is usually used in
defining large $N$ matrix models instead of the adjoint
representation.
However, the similar large $N$ limit of the $N\times N$ matrices
of the $SU(N)$ generators are divergent, while the matrix elements of
the generators of the Poisson algebra with the same basis are finite.
These points are discussed in section \ref{sec:aadag}.
The last section is devoted to summary and discussion.

\section{Representation with functions on world-sheet}
Lie algebra of infinitesimal area-preserving diffeomorphism is
represented by the operators over functions on a surface, which we
call world-sheet,
\begin{equation}
 X_{f(\sigma)}\equiv\omega^{ab}\partial_af(\sigma)\partial_b,
	\label{eq:Poisson-dif}
\end{equation}
where $\omega$ is the symplectic two-form of the world-sheet
and it satisfies
\begin{equation}
  \omega^{ad}\partial_d\omega^{bc} +
  \omega^{bd}\partial_d\omega^{ca} +
  \omega^{cd}\partial_d\omega^{ab} =0\,.
\end{equation}
Actually we can straightforwardly show the following relation,
\begin{equation}
 [X_{f(\sigma)},X_{g(\sigma)}]
  =X_{f(\sigma)}X_{g(\sigma)}-X_{g(\sigma)}X_{f(\sigma)}
  =X_{\{f(\sigma),g(\sigma)\}}\,, \label{eq:ComX}
\end{equation}
where $\{f(\sigma),g(\sigma)\}$ is the Poisson bracket defined by
\begin{equation}
 \{f(\sigma),g(\sigma)\} \equiv
	\omega^{ab}\partial_a f(\sigma) \cdot \partial_b g(\sigma)\,.
\end{equation}
Let us consider the case that the world-sheet is a two-dimensional
torus, $\sigma\in [\,0,1)\times [\,0,1)$, and hence
$\omega^{ab}=\epsilon^{ab}$.
Functions on the torus can be expanded in Fourier series and the
operator (\ref{eq:Poisson-dif}) is represented with the basis as
\begin{equation}
  \left(\,X_{\bf m}\,\right)_{\bf kl} \equiv \int d^2\sigma\,
   e^{-2\pi i {\bf k}\cdot \sigma}\, X_{e^{2\pi i {\bf m}\sigma}}\,
   e^{2\pi i {\bf l}\cdot\sigma}= -4\pi^2 {\bf m}\times{\bf l}\,
   \delta_{{\bf m}\,({\bf k-l})}\,,\label{eq:Padj}
\end{equation}
where ${\bf m}\times{\bf n} \equiv\epsilon^{ab} m_a n_b$.
Then eq.(\ref{eq:ComX}) is represented by
\begin{equation}
 (\,[X_{\bf m},X_{\bf n}]\,)_{\bf kl}=
 \sum_{\bf j} \left[
  \left(\,X_{\bf m}\,\right)_{\bf kj}\,
    \left(\,X_{\bf n}\,\right)_{\bf jl}
  - \left(\,X_{\bf n}\,\right)_{\bf kj}\,
    \left(\,X_{\bf m}\,\right)_{\bf jl}\right]
  = -4\pi^2 {\bf m}\times{\bf n}\,
	\left(\,X_{\bf m+n}\,\right)_{\bf kl}\,.\label{eq:PadjM}
\end{equation}
This means that $(\,X_{\bf m}\,)_{\bf kl}$ in (\ref{eq:Padj}) are the
matrix elements for the adjoint representation of the Poisson algebra.
It seems natural to use fields over the world-sheet with the Poisson
bracket instead of the operators (\ref{eq:Poisson-dif}) as
representation of the Poisson algebra, however, the fields on the
world-sheet themselves cannot represent a matrix algebra.
On the other hand, with the operators (\ref{eq:Poisson-dif}) we have
the Poisson algebra in (\ref{eq:ComX}) and the matrix algebra
(\ref{eq:PadjM}) as well.
Since the fields on the world-sheet have one-to-one correspondence (up
to constant functions) with the operators (\ref{eq:Poisson-dif}), we
can take either of them to consider the representation of the Poisson
algebra.
Our aim is to study the large $N$ limit of matrix models, and then it
is more useful to take those operators.

Next we consider the matrix representation of $SU(N)$ algebra in the
adjoint representation and its large  $N$ limit.
To construct the representation, however, let us first consider
$SU(N)$ algebra in the fundamental representation.
We can choose the generators of $SU(N)$ in the following form
\cite{KU,bars,zachos,NBIs},\footnote{We assume that $N$ is odd for
definiteness. Transformation to the Cartan-Weyl basis is shown in
\cite{zachos}.}
\begin{equation}
 Y^{SU(N)}_{(m_1,m_2)}
  =e^{i{\frac{2\pi}{N}}m_1m_2}\,U^{m_1}\,V^{m_2},\label{eq:SU_Ngen}
\end{equation}
where $U$ and $V$ are $N\times N$ clock and shift matrices,
respectively,
\begin{equation}
 U=\left(
    \begin{array}{@{\,}ccccc@{\,}}
     1&&&&\Bzero\\
     &e^{i\frac{4\pi}{N}}&&&\\
     &&\ddots&&\\
     \multicolumn{2}{c}{\mbox{\bg 0}}&&&e^{{i\frac{4\pi}{N}}(N-1)}\\
    \end{array}\right),~~~
 V=\left(
    \begin{array}{@{\,}ccccc@{\,}}
     0&1&&&\Bzero\\
     &0&1&&\\
     &&\ddots&\ddots&\\
     &\mbox{\bg 0}&&0&1\\
     1&&&&0
    \end{array}\right)\,, \label{eq:ClockAndShift}
\end{equation}
which satisfy $U^N=V^N= 1$ and hence $Y^{SU(N)}_{(m_1+k_1N,m_2+k_2N)}
= Y^{SU(N)}_{(m_1,m_2)} \hspace{1ex}(k_i\in\Integer)$.
The commutation relations are given by
\begin{equation}
 [\,Y^{SU(N)}_{\bf m},Y^{SU(N)}_{\bf n}\,]
  =-2i\sin\left(\frac{2\pi}{N}{\bf m\times n}\right)
  Y^{SU(N)}_{\bf m+n}. \label{eq:ComSU_N}
\end{equation}
Thus the matrix elements of $SU(N)$ generators in the adjoint
representation can be taken as\footnote{We have, of course,
hermitian combinations, $Z^N_{\bf m}+Z^{N}_{-\bf m}$ and $i(Z^N_{\bf
m}-Z^N_{-\bf m})$, so that $Z^N_{\bf m}$ are the generators of
$SU(N)$.}
\begin{equation}
 (Z^{N}_{\bf m})_{\bf kl}
  \equiv -2i\sin\left(\frac{2\pi}{N}{\bf m\times l}\right)
  \delta_{\bf m (k-l)}\,.  \label{eq:adSU_N}
\end{equation}
Actually, we have
\begin{eqnarray}
 \left(\,[ Z^N_{\bf m},Z^N_{\bf m} ]\,\right)_{\bf k l} &=&
 \sum_{\bf i}
  \left[ \left(Z^{N}_{\bf m}\right)_{\bf k i}\,
	\left(Z^{N}_{\bf n}\right)_{\bf i l}
	- \left(Z^{N}_{\bf n}\right)_{\bf k i}\,
	\left(Z^{N}_{\bf m}\right)_{\bf i l}\right] \nn
  &=&  -2i \sin\left( \frac{2\pi}{N} {\bf m}\times{\bf n}\right)
  \,\left(Z^{N}_{\bf m+n}\right)_{\bf k l}\,,
\end{eqnarray}
that is, the generators of $SU(N)$ are realized by the matrices
$(Z^{N}_{\bf n})_{\bf k l}$.

Now we consider the following large $N$ limit of the algebra.
In fact, multiplying eq.(\ref{eq:ComSU_N}) by $N^2$, we have
\begin{equation}
 [\,N\,Y^{SU(N)}_{\bf m}, N\,Y^{SU(N)}_{\bf n}\,]
  =-2i \left\{ N\sin\left(\frac{2\pi}{N}{\bf m\times n}\right)\right\}
  N\,Y^{SU(N)}_{\bf m+n}. \label{eq:NComSU_N}
\end{equation}
Since the structure constant,
$N\sin\left(2\pi{\bf m\times n}/N\right)$, becomes
$2\pi{\bf m\times n}$ for fixed ${\bf m}$ and ${\bf n}$ when
$N\to\infty$, one may naively expect that $X_{\bf m} = 
\lim_{N\to\infty}[N\,Y^{SU(N)}_{\bf m}]$ will satisfy $[X_{\bf
m},X_{\bf n}]= -4\pi i\,{\bf m\times n}\,X_{\bf m+n}$, however, one
should be careful with the limit in the defining equation of
$X_{\bf m}$, that is, $N\,Y^{SU(N)}_{\bf m}$ could diverge and hence
the limit does not exist.
Let us consider eq.(\ref{eq:adSU_N}), the $SU(N)$ generators in the
adjoint representation.
In the large $N$ limit, they agree with those of the Poisson algebra
in eq.(\ref{eq:Padj}) when we rescale $(Z^{N}_{\bf m})_{\bf kl}$
as\footnote{Notice that for fixed $k,l$ and ${\bf m}$, the matrix
element of $N\left(Z^{N}_{\bf m}\right)_{\bf kl}$ is finite in
the $N\to\infty$ limit. This is not the case in the fundamental
representation, as will be seen in the next section.}
\begin{equation}
 i\pi N(Z^{N}_{\bf m})_{\bf kl}\,
  \mathop{\longrightarrow}_{N\rightarrow\infty}\,
  (X_{\bf m})_{\bf kl}\,. \label{eq:ad-infty}
\end{equation}
We shall give the operators with functions on the torus whose matrix
representation coincides with eq.(\ref{eq:adSU_N}).
They are given by
\begin{equation}
  Z^{N}_f(\sigma) = f(\sigma) \left(
    \exp\left[-\frac{i}{2\pi N}\,{\bm\partial}
	 (\ln{f})\times {\bm\partial}\right]
    - \exp\left[\frac{i}{2\pi N}\,{\bm\partial}
    (\ln{f})\times {\bm\partial}\right]\right)\,.
\end{equation}
In fact, the matrix elements of the operators with
$f(\sigma)=f_{\bf m}(\sigma)\,(\equiv\exp(i2\pi{\bf m}\cdot\sigma)\,)$
are given by
\begin{equation}
 (Z^{N}_{\bf m})_{\bf k l} \equiv
  \int d^2\sigma f^*_{\bf k}(\sigma)\,Z^{N}_{\bf m}(\sigma)\,
  f_{\bf l}(\sigma)
  =-2i\sin\left(\frac{2\pi}{N}{\bf m\times l}\right)\delta_{{\bf
  m}\,({\bf k-l})}\,. \label{eq:AdMatrix}
\end{equation}
Furthermore, we can straightforwardly take $N\rightarrow\infty$ limit
of $N Z^{N}_f(\sigma)$ as
\begin{eqnarray}
  N Z^{N}_f(\sigma) &= &N f(\sigma) \left(
    \exp\left[-\frac{i}{2\pi N}\,{\bm\partial}
	 (\ln{f})\times {\bm\partial}\right]
    - \exp\left[\frac{i}{2\pi N}\,{\bm\partial}
    (\ln{f})\times {\bm\partial}\right]\right)\nn
  &=& -N f \left( \frac{i}{\pi N}{\bm\partial}
	 (\ln{f})\times {\bm\partial} +
    O(\frac{1}{N^2})\right)\nn
  &\rightarrow& -\frac{i}{\pi}{\bm\partial} f\times {\bm\partial}\,,
\end{eqnarray}
which coincides with $X_f$.

We realized $SU(N)$ algebra in the infinite dimensional operator space
and we can realize the finite dimensional representation of those
operators by restricting fields due to the periodicity in
eq.(\ref{eq:adSU_N}).
Since $(Z^{N}_{\bf m})_{\bf k l}$ in eq.(\ref{eq:AdMatrix}) are
$(N^2-1)\times (N^2-1)$ matrices, we can hardly regard that this
representation is the large $N$ limit of the $N\times N$ matrix
representation of $SU(N)$ algebra. So we need other ways to represent
$SU(N)$ and the Poisson algebra simultaneously to formulate the large
$N$ limit of matrix models.

\section{Representation with \boldmath{$a$} and \boldmath{$a^\dagger$}
\label{sec:aadag}}
We need the fundamental representation of the Poisson algebra, which
could be regarded as the large $N$ limit of $SU(N)$ algebra, to
construct the large $N$ matrix model.
It is natural to think that the representation is embedded in an
infinite dimensional representation, and we could relate the large $N$
limit of $SU(N)$ and the Poisson algebra as in the adjoint
representation case in eq.(\ref{eq:ad-infty}).
As we show below, these algebras can be embedded in the operator space
generated by the creation and annihilation operators and its
representation space,
\begin{equation}
 [\,a,a^\dagger\,]=1,~~a\ket{0}=0,~~
  \ket{n}\equiv\frac{1}{\sqrt{n!}}(a^\dagger)^n\ket{0}\,.
  \label{eq:FockSpace}
\end{equation}
Since the commutator of $a$ and $a^\dagger$ is the same as
the one between the coordinates in a non-commutative space, functions
of $a$ are regarded as those on a non-commutative space.

We show how the generators of the algebras for the fundamental
representation are embedded in the operator space.
First, let us consider the following operators \cite{hoppesch},
\begin{equation}
 T^{N,\eta}_{\bf m}
  =  \exp\left[2i\sqrt{\frac{\pi}{N}} \left(\eta m_1 a + i
  \frac{m_2}{\eta} a^\dagger\right)\right],   \label{eq:TSU(N)}
\end{equation}
where $\eta$ is an arbitrary parameter.
The commutators  of $T^{N,\eta}_{\bf m}$ are easily calculated as
\begin{equation}
 [\,T^{N,\eta}_{\bf m}, T^{N,\eta}_{\bf n}\,]
  =-2i\sin\left(\frac{2\pi}{N}{\bf m\times n}\right)
	 T^{N,\eta}_{\bf m+n}\,, \label{eq:TSU(N)Com}
\end{equation}
which are certainly commutation relations for $SU(N)$ algebra as in
eq.(\ref{eq:ComSU_N}), although $T^{N,\eta}_{\bf m}$ do not satisfy
the periodicity,
$T^{N,\eta}_{{\bf m}+N{\bf k}} \neq T^{N,\eta}_{\bf m}$.

We construct $N\times N$ dimensional representation of
(\ref{eq:TSU(N)}) with the following bases of the vector space
and its dual space, respectively,
\begin{eqnarray}
 \Nket{k;\eta}&\equiv&
  \exp\left[2\sqrt\frac{\pi}{N}\,
  \frac{k a^\dagger}{\eta}\right]\ket{0}\,,\label{eq:ND-Ket}\\
 \Nbra{k;\eta}&\equiv&
  \frac{1}{N}\bra{0}\sum_{m=0}^{N-1}
  \exp\left[-i\frac{4\pi}{N}km\right]\,
  \exp\left[2i\sqrt\frac{\pi}{N}\,\eta ma\right]\,,
  \label{eq:ND-Bra}
\end{eqnarray}
where $k=0,\cdots,N-1$. The inner products are
$\Nbracket{k;\eta}{l;\eta}=\delta_{k\,l}$.
Thus the following operator,
\begin{equation}
 P^{\eta}_{N}\equiv\sum_{k=0}^{N-1} \Nket{k;\eta}\Nbra{k;\eta},
  \label{eq:projection}
\end{equation}
is a projection operator to a $N$ dimensional Fock subspace.
Then the matrix elements of $T^N_{\bf m}$ are given by
\begin{eqnarray}
 \Nbra{k;\eta} T^{N,\eta}_{\bf m}\Nket{l;\eta}&=&
  \frac{1}{N}\sum_{n=0}^{N-1}\,e^{-i\frac{2\pi}{N}m_1m_2}
	 e^{4\pi i\frac{n(l-k)}{N}}\,e^{-4\pi i\left(\frac{n\eta
  m_2}{\eta N} - \frac{\eta m_1 l}{\eta}\right)}\nn
  &=&e^{-i\frac{2\pi}{N}m_1m_2}
  e^{-i\frac{4\pi}{N}m_1k}\delta_{l\,[k+m_2]_N},
  \label{eq:SU(N)Mx-ele}
\end{eqnarray}
where we have introduced the notation that $[k]_{N} \equiv
k~\mbox{mod}~N$ ,i.e., $0\leq [k]_N< N$.
Note that the above matrix elements are exactly the same as those of
the generators of $SU(N)$ in eq.(\ref{eq:SU_Ngen}) with the clock and
the shift matrices in eq.(\ref{eq:ClockAndShift}) and the periodicity
is realized on the representation,  $\Nbra{k;\eta}
T^{N,\eta}_{(m_1+Nk_1,m_2+Nk_2)}\Nket{l;\eta} = \Nbra{k;\eta}
T^{N,\eta}_{\bf m}\Nket{l;\eta}$.
This is a $N\times N$ matrix representation of $SU(N)$ algebra, so
that this operator space will be suitable to formulate the large $N$
limit of matrix models.

On the other hand, the generators of the Poisson algebra are
constructed as
\begin{equation}
 T^{(P,\xi)}_{\bf m}
  =\left\{1-2i\sqrt{\pi}
	\left(\xi m_1a+i\frac{m_2}{\xi}a^\dagger\right)\right\}\,
	\exp\left[2i\sqrt{\pi}
	\left(\xi m_1a+i\frac{m_2}{\xi}a^\dagger\right)\right],
  \label{eq:TPoisson}
\end{equation}
where $m_i,n_i \in \Integer$ and $\xi$ is an arbitrary constant.
In fact, their commutation relations are
\begin{equation}
 [\, T^{(P,\xi)}_{\bf m},T^{(P,\xi)}_{\bf n}\,]
  = 4\pi i\,{\bf m\times n}\,T^{(P,\xi)}_{\bf m+n}\,,
  \label{eq:ComTPoisson}
\end{equation}
whose proof are given in the appendix.
It is natural to expect that we can construct the fundamental
representation of the Poisson algebra with suitable bases of a
vector space and its dual space as in $SU(N)$ case in
eqs.(\ref{eq:ND-Ket}) and (\ref{eq:ND-Bra}), and the large $N$ limits
of the $SU(N)$ generators in the fundamental representation
coincide with those of the Poisson algebra as in the adjoint
representation case in eq.(\ref{eq:ad-infty}) due to
ref.\cite{zunger}. However, this seems to be incorrect.
In the above arguments, we expect implicitly that the large $N$ limit
of the vector space spanned by $\Nket{k;\eta}$ in eq.(\ref{eq:ND-Ket})
and its dual space in eq.(\ref{eq:ND-Bra}) coincide with the whole
Fock space and its dual, respectively, which means that the large $N$
limit of the projection operator $P^\eta_N$ (\ref{eq:projection}) is
the identity on the Fock space. However, we can see that
$\mathop{\lim}\limits_{N\rightarrow\infty}\bra{0}\,[a,P^\eta_N]\,
\ket{1}$ is divergent by elementary calculations. This means that
since $\mathop{\lim}\limits_{N\rightarrow\infty}P^\eta_N$, if it
exists, is the projection operator, $P^\eta_N$ cannot be an identity
operator even in $N\rightarrow\infty$ and the large $N$ limit of the
vector space (or its dual space) is not the whole Fock space but its
subspace. This fact implies that the generators of the Poisson algebra
in eq.(\ref{eq:TPoisson}) cannot be represented with the vector space
in eq.(\ref{eq:ND-Ket}) and its dual space in eq.(\ref{eq:ND-Bra}),
or we need the whole Fock space to represent the Poisson algebra.
The matrix elements $\bra{k} T^{(P,\xi)}_{\bf m} \ket{l}\quad
(k,l=0,1,\cdots)$ can be straightforwardly calculated as\footnote{
For fixed $k,l$ and ${\bf m}$, the matrix elements are finite.}
\begin{eqnarray}
 &&\hspace{-3ex}\bra{k} T^{(P,\xi)}_{\bf m} \ket{l}\nn
 &&=\left\{\renewcommand{\arraystretch}{2}\begin{array}{@{\,}ll}
   \displaystyle \sqrt{\frac{l!}{k!}}\,\lambda_2^{k-l}
	\left[(1-z) L_l^{(k-l)}(z) + z L_l^{(k-l+1)}(z)
	- (l+1)	L_{l+1}^{(k-l-1)}(z)   \right]\,, & (k>l)\\
   \displaystyle (1-z)L_k^{(0)}(z) + 2z L_{k+1}^{(0)}(z)\,,&(k=l)\\
   \displaystyle \sqrt{\frac{k!}{l!}}\,\lambda_1^{l-k}
	\left[(1-z) L_k^{(l-k)}(z) + z L_k^{(l-k+1)}(z)
	- (k+1) L_{k+1}^{(l-k-1)}(z)   \right]\,, & (k<l)
 \end{array}\right.
\end{eqnarray}
where
\begin{equation}
	z= 4\pi i m_1 m_2\,,\hspace{3ex}
	\lambda_1 = 2i\sqrt{\pi} m_1\xi\,,\hspace{3ex}
	\lambda_2 = - \frac{2\sqrt{\pi} m_2}{\xi}\,,
\end{equation}
and $L_n^{(\alpha)}(x)$ are the generalized Laguerre polynomials.
The commutation relations of the Poisson algebra are expressed with
the matrix elements\footnote{For fixed $k,l,{\bf m}$ and ${\bf n}$,
the summation over $p$ is well-defined.},
\begin{equation}
 \sum_{p=0}^{\infty} \Big[ \bra{k}T^{(P,\xi)}_{\bf m}\ket{p}\,
	\bra{p} T^{(P,\xi)}_{\bf n}\ket{l} -
    \bra{k} T^{(P,\xi)}_{\bf n}\ket{p}\,
	\bra{p} T^{(P,\xi)}_{\bf m}\ket{l}\Big]
    =  4\pi i\,{\bf m\times n}\,\bra{k} T^{(P,\xi)}_{\bf m+n} \ket{l}\,.
\end{equation}

Using these coefficients $\bra{k}T^{(P,\xi)}_{\bf m}\ket{l} \equiv
W_{\bf m}^{kl}$, we can give another expression of the operators for
the Poisson algebra,
\begin{eqnarray}
  V_{\bf m} &\equiv& \sum_{k,l=0}^{\infty}
	W_{\bf m}^{kl}\,\ket{k}\bra{l}\nn
  &=&  \sum_{k,l}\,\frac{W_{\bf m}^{kl}}{\sqrt{k!\,l!}}\,
	:(a^\dagger)^k\,e^{-a^\dagger a}\,a^l: \,, \label{eq:V_m}
\end{eqnarray}
where ``$:\cdot :$'' stands for the normal product.
Furthermore, using the operator $\ket{k}\bra{l}$ we can construct the
generators of $SU(N)$ as
\begin{eqnarray}
  V_{\bf m}^{SU(N)} &=& e^{\frac{2\pi i}{N}m_1m_2}\left[
	\sum_{k=0}^{N-1-m_2}\,e^{\frac{4\pi i}{N}km_1}\,
	\ket{k}\bra{k\!+\!m_2}
    + \sum_{k=N-m_2}^{N-1}\,e^{\frac{4\pi i}{N}km_1}\,
	\ket{k}\bra{k\!+\!m_2\!-\!N}\,\right]\nn
  &=& e^{\frac{2\pi i}{N}m_1m_2}\sum_{k=0}^{N-1}\,
	e^{\frac{4\pi i}{N}km_1}\, :(a^\dagger)^k\,e^{-a^\dagger a}
	\, a^{[k+m_2]_N}:\,,
\end{eqnarray}
We can easily see that these $V_{\bf m}^{SU(N)}$ have the periodicity,
$V_{(m_1+Nk_1,m_2+Nk_2)}^{SU(N)}= V_{(m_1,m_2)}^{SU(N)}$, and they
satisfy the same commutation relation in eq.(\ref{eq:TSU(N)Com}).
Hence $V_{\bf m}^{SU(N)}$ have the $N\times N$ matrix representation
in the $N$-dimensional vector space with basis
$\{\ket{0},\ket{1},\cdots,\ket{N\!-\!1}\}$ and its dual with
$\{\bra{0},\bra{1},\cdots,\bra{N\!-\!1}\}$,
\begin{eqnarray}
  (\,V_{\bf m}^{SU(N)}\,)_{kl} &=& \bra{k}V_{\bf m}^{SU(N)}\ket{l}
	\nn
  &=& e^{-i\frac{2\pi}{N}m_1m_2}\,e^{-i\frac{4\pi}{N}m_1k}\,
	\delta_{l[k+m_2]_N}
 \Big(= \Nbra{k;\eta} T^{N,\eta}_{\bf m}\Nket{l;\eta}\Big)
	\,.\label{eq:SuNmtrxele}
\end{eqnarray}
However, we shall see that the $N\times N$ matrices, $N \left( V_{\bf
m}^{SU(N)} \right)$, do not have the well-defined $N\rightarrow\infty$
limit\footnote{Even for fixed $k,l$ and ${\bf m}$, some matrix
elements are divergent when $N\to\infty$.} and hence, contrary
to eq.(\ref{eq:ad-infty}), $N \left(V_{\bf m}^{SU(N)}\right)$ cannot
become $\Big(V_{\bf m}\Big)$ in the large $N$ limit.

\section{Summary and discussion}
We have studied whether the large $N$ limit of $SU(N)$ algebra
coincide with the Poisson algebra. In the adjoint representation, we
have shown their coincidence by comparing their matrix elements as in
eq.(\ref{eq:ad-infty}), while they do not coincide with each other in
the fundamental representation.
In fact, the matrix elements of the $SU(N)$ generators can be written
by eq.(\ref{eq:SuNmtrxele}) (or eq.(\ref{eq:SU(N)Mx-ele})) and the
rescaled matrix which is multiplied by $\mathcal{O}(N)$
constant (cf. eq.(\ref{eq:ad-infty})), is divergent in the
$N\rightarrow\infty$ limit. In other words, we cannot give a sequence 
of $N\times N$ matrix representation of $SU(N)$ algebra with the
structure constant which is proportional to $N\sin(2\pi{\bf
m}\times{\bf n}/N)$ that goes to the representation of the Poisson
algebra.

Let us consider $N^2$ operators, $V_{\bf m}^N \equiv\sum_{k,l=0}^{N-1}
W_{\bf m}^{kl}\,\ket{k}\bra{l}\quad (0\leq |m_a|\leq (N-1)/2)$,
which give another regularization of the Poisson algebra. They go to
$V_{\bf m}$  in the $N\rightarrow\infty$ limit, however, we can see
that $V_{\bf m}^N$ are linear dependent and hence $V_{\bf m}^N$ (or
$V_{\bf m}^N/N$) do not satisfy $SU(N)$ algebra.
On the other hand, another set of $N^2$ operators,  $V_{\bf m}^N
\quad(1\leq \pm m_a\leq (N\pm1)/2)$, seems to be linear
independent\footnote{We have checked linear independence for small
$N$'s but we have not given a general proof.} and they satisfy
the $SU(N)$, actually $U(N)$, algebra whose structure constants are
different from those in eq.(\ref{eq:TSU(N)Com}). For finite $N$,
$V_{(m_1,0)}^N$ and $V_{(0,m_2)}^N$ are given by the linear
combinations of $V_{\bf m}^N\quad(1\leq\pm m_a\leq (N\pm 1)/2)$,
respectively, while all $V_{\bf m}$ are independent.
Then, a sequence of the set of $(N+1)^2$ operators,
$V_{\bf m}^N\quad (0\leq\pm m_a\leq (N\pm 1)/2)$, which
are linear dependent for finite $N$, will go to $V_{\bf m}$ in the
$N\rightarrow\infty$ limit.
This implies that the $N\times N$ matrix models may not be suitable to
regularize theories with area-preserving diffeomorphisms\footnote{
In this case the Poisson algebra is realized with the creation
and the annihilation operators. Then the arguments in
ref.\cite{zunger} will be inapplicable since the generators in
eq.(\ref{eq:TPoisson}) or eq.(\ref{eq:V_m}) are {\it not} associated
with functions on any ordinary compact world-sheet.}.
How the dependent set of matrices becomes independent ones, or
the algebra for the $(N+1)^2$ matrices, is deserved to be investigated
further to understand the matrix models.
\section*{acknowledgments}
We would like to thank Akihiro Tsuchiya and Hiroaki Kanno for
discussions. The work of SU is supported in part by the Grant-in-Aid
for Scientific Research No.13135212.

\appendix
\section{Calculation of eq.(\ref{eq:ComTPoisson})}
Once we first notice the following identities,
eq.(\ref{eq:ComTPoisson}) can be shown straightforwardly,
\begin{eqnarray}
e^{2i\sqrt{\pi}
	\left(\xi m_1a+i\frac{m_2}{\xi}a^\dagger\right)}\,
 e^{2i\sqrt{\pi}
	\left(\xi n_1a+i\frac{n_2}{\xi}a^\dagger\right)}
 =e^{2i\sqrt{\pi}
	\left[\xi
	(m_1+n_1)a+i\frac{m_2+n_2}{\xi}a^\dagger\right]}.
	\label{eq:TtimesT}
\end{eqnarray}
Actually we have
\begin{equation}
 [\,2i\sqrt{\pi}\,(\xi m_1a+i\frac{m_2}{\xi}a^\dagger),\,
  2i\sqrt{\pi}\,(\xi n_1a+i\frac{n_2}{\xi}a^\dagger)\,]
  =-4\pi i\,(m_1n_2-m_2n_1)\in 4\pi i\,{\mathbb Z}, \label{eq:com-a}
\end{equation}
so that the phase factor in eq.(\ref{eq:TtimesT}), which comes from
eq.(\ref{eq:com-a}), are trivial.
To complete the calculation, we must evaluate cross terms as
\begin{eqnarray}
[\,2i\sqrt{\pi}\,(\xi m_1a+i\frac{m_2}{\xi}a^\dagger) ,\,
    e^{2i\sqrt{\pi}\left(\xi
    n_1a+i\frac{n_2}{\xi}a^\dagger\right)}\,]=
  -4\pi i\,(m_1n_2-m_2n_1)\,
  e^{2i\sqrt{\pi}
	 \left(\xi n_1a+i\frac{n_2}{\xi}a^\dagger\right)}.
\label{eq:com-diff}
\end{eqnarray}
Then eqs.(\ref{eq:TtimesT}, \ref{eq:com-a}, \ref{eq:com-diff}) lead to
eq.(\ref{eq:ComTPoisson}).

\end{document}